# Slow Auger Recombination of Charged Excitons in Nonblinking Perovskite Nanocrystals without Spectral Diffusion

Fengrui Hu,[1] Chunyang Yin,[1] Huichao Zhang,[1,3] Chun Sun,[2] William W. Yu,[2] Chunfeng Zhang,[1] Xiaoyong Wang,[1,*] Yu Zhang,[2,*] and Min Xiao[1,4,*]

[1] *National Laboratory of Solid State Microstructures, School of Physics, and Collaborative Innovation Center of Advanced Microstructures, Nanjing University, Nanjing 210093, China*
[2] *State Key Laboratory on Integrated Optoelectronics and College of Electronic Science and Engineering, Jilin University, Changchun 130012, China*
[3] *College of Electronics and Information, Hangzhou Dianzi University, Xiasha Campus, Hangzhou 310018, China*
[4] *Department of Physics, University of Arkansas, Fayetteville, Arkansas 72701, USA*

**Abstract:** Over the last two decades, intensive research efforts have been devoted to the suppressions of photoluminescence (PL) blinking and Auger recombination in metal-chalcogenide nanocrystals (NCs), with significant progresses being made only very recently in several specific heterostructures. Here we show that nonblinking PL is readily available in the newly-synthesized perovskite $CsPbI_3$ (cesium lead iodide) NCs, and their Auger recombination of charged excitons is greatly slowed down, as signified by a PL lifetime about twice shorter than that of neutral excitons. Moreover, spectral diffusion is completely absent in single $CsPbI_3$ NCs at the cryogenic temperature, leading to a resolution-limited PL linewidth of ~200 μeV.

Auger recombination describes an intriguing optoelectronic phenomenon in semiconductor nanocrystals (NCs) whereby the exciton energy is nonradiatively transferred to an extra charge instead of being converted to a photon [1]. For regular metal-chalcogenide NCs (*e.g.*, CdSe), the Auger processes of multi-excitons and charged excitons normally occur at the sub-nanosecond timescale [2], much shorter than the radiative lifetime of tens of nanoseconds for single neutral excitons [3]. This limited survival time of multi-excitons in semiconductor NCs makes it difficult to achieve long optical gain in lasers [4] as well as to obtain high power conversion efficiency in photodetectors [5] and solar cells [6,7] based on the carrier multiplication effect [8]. Meanwhile, multi-exciton ionization [9] and direct carrier trapping [10] are the two main channels for the intermittent formation of charged excitons in a single NC, resulting in the photoluminescence (PL) blinking "off" periods [11] that are detrimental to the single-photon emitting characteristics [12] and the brightness of light-emitting diodes [13] normally operated with imbalanced injection of charge carriers. So far, effective suppressions of both Auger recombination and PL blinking has been realized only in several specific heterostructures such as "giant" CdSe/CdS core/shell NCs [14,15]. Besides PL blinking, another notorious optical property manifested at the single-NC level is the random shift of the PL peak energy that is denoted as the spectral diffusion effect [16]. Although a PL linewidth as narrow as hundreds of μeV can be occasionally observed from single CdSe NCs [17], its stability over long measurement times is still being actively pursued in order to elongate the exciton dephasing time for the implementation of coherent optical measurements [18,19].

Semiconductor perovskite NCs of cesium lead halide have been synthesized very recently [20] as a novel fluorescent nanostructure capable of emitting single photons with a large absorption cross section and a short radiative lifetime [21-23]. Here we show that, at room temperature and with low-power laser excitation, nonblinking PL is easily achieved in single perovskite CsPbI$_3$ (cesium lead iodide) NCs synthesized from a facile colloidal approach. With high-power laser excitation, PL blinking is triggered in single CsPbI$_3$ NCs by the photo-ionization effect that creates two types of charged excitons with opposite signs. The "grey" intensity level in the PL blinking time trace is related to the charged exciton with Auger-mediated weak fluorescence; Auger recombination in the other type of charged exciton is nearly eliminated so that its fluorescent photons contribute to the blinking "on" intensity level with a PL lifetime almost twice shorter than that of the neutral exciton. At cryogenic temperature and with low-power laser excitation, a resolution-limited PL linewidth of ~200 μeV is measured for single CsPbI$_3$ NCs without the influence of the spectral diffusion effect.

According to a previous report [20], the CsPbI$_3$ NCs used in our experiment were synthesized (see Supplemental Material [24] for experimental details) with an average size of ~9.3 nm (see Fig. S1 of the Supplemental Material [24]) and a PL peak around ~690 nm at room temperature (see Fig. S2 of the Supplemental Material [24]). Single CsPbI$_3$ NCs placed on top of a silica substrate were excited at ~570 nm with a picosecond pulsed laser and the PL signals were detected by a time-correlated single-photon counting system (see Supplemental Material [24] for experimental details). In Fig. 1(a), we plot the PL intensity time trace of a nonblinking CsPbI$_3$ NC excited at <N> = ~0.03, where <N> represents the average number of photons absorbed per NC per pulse (see Fig. S3 of the Supplemental Material [24]). The PL decay curve of this single CsPbI$_3$ NC is plotted in Fig. 1(b), which can be well fitted by a single-exponential function with a radiative lifetime of ~43.4 ns for single neutral excitons (Xs). Single-photon emission of this single CsPbI$_3$ NC was confirmed in Fig. 1(c) from the second-order autocorrelation function $g^{(2)}(\tau)$ measurement, where the average area ratio between the central and the side peaks was calculated to be ~0.05. Nonblinking PL was observed in nearly ~50 of the ~60 single $CsPbI_3$ NCs excited in our experiment at $<N>$ = ~0.03, which can be attributed mainly to the nonexistence of defect traps accessible to the band-edge charge carriers so that the formation of charged excitons is prohibited. For each of the other ~10 single $CsPbI_3$ NCs, there does exist defect traps to capture either the band-edge electron or hole, leading to occasional switching of the PL intensity from the "on" to the lower intensity levels due to nonradiative Auger recombination of charged excitons (see Fig. S4 of the Supplemental Material [24] ).

Although the PL blinking effect was completely suppressed in most of the single $CsPbI_3$ NCs excited at low power, it was triggered in all of the single $CsPbI_3$ NCs studied in our experiment with high-power excitation due to the formation of charged excitons from the multi-exciton ionization process [9,25]. In Fig. 2(a), we plot the nonblinking PL time trace of a single $CsPbI_3$ NC excited at $<N>$ = ~0.02, whose PL intensity distribution is shown in Fig. 2(b) with a single "on"-level peak. The blinking PL time trace of the same single NC excited at $<N>$ = ~1.7 is plotted in Fig. 2(c), where a "grey" intensity level can be clearly resolved in addition to the normal "on" and "off" ones. As can be seen in Fig. S5 of the Supplemental Material [24] from an enlarged part of this PL time trace marked by the red box and in Fig. 2(d) from the PL intensity distribution, the "grey" intensity level is associated with a PL efficiency of ~15% relative to that of the "on" intensity level. This kind of "grey" intensity level was previously reported in single metal-chalcogenide CdSe NCs as a sign of slightly-reduced Auger recombination in charged excitons [25-30]. When a single $CsPbI_3$ NC is excited at $<N>$ = ~1.7 to create mainly Xs and neutral biexcitons (XXs), we assume for convenience that it is the electron in the X being pumped to a higher excited state after receiving the XX recombination energy in an Auger process. The probability for this electron to be captured by an external trap is higher than would it stay in the band-edge state [31], thus leaving an unpaired hole in the NC. A positively-charged two-exciton state would be prepared in the next excitation event so that PL photons of the "grey" intensity level should be mainly contributed by positively-charged single excitons ($X^+$s) when assuming that its PL efficiency is significantly higher than that of positively-charged biexcitons.

The "on"-level PL decay curve of the CsPbI$_3$ NC excited at $<N>$ = ~0.02 is plotted in Fig. 2(e), which can be well fitted by a single-exponential function with a radiative lifetime of ~55.3 ns for Xs (see fitting residual in Fig. 2(f)). At $<N>$ = ~1.7, the "on"-level PL decay curve shown in Fig. 2(e) can only be roughly fitted by a single-exponential function with a shortened PL lifetime of ~42.0 ns (see fitting residual in Fig. 2(f)), which is unexpected for the radiative lifetime of Xs since it should be independent of the laser excitation power. However, the same PL decay curve can be well fitted by a double-exponential function of $A_1 e^{-t/\tau_1} + A_2 e^{-t/\tau_2}$ (see fitting residual in Fig. 2(f)), with $A_1$ ($A_2$) and $\tau_1$ ($\tau_2$) being the amplitude and the value of the slow (fast) lifetime component, respectively. The slow lifetime ($\tau_1$) of ~51.0 ns is close to the radiative lifetime of ~55.3 ns measured at $<N>$ = ~0.02 for Xs, while the fast lifetime ($\tau_2$) of ~33.4 ns could be related to negatively-charged excitons ($X^-$s) with strongly-reduced Auger recombination that was previously observed in single "giant" CdSe/CdS NCs [32,33]. In contrast to the "grey"-level case, the "on"-level $X^-$ should be formed after a photo-excited hole is ejected out of the NC in a XX Auger ionization process. The ratio of $\tau_1/\tau_2$ was calculated to be ~1.53 for this specific NC and ~1.83 on average for the ~20 NCs studied in our experiment (see Fig. S6 of the Supplemental Material [24]), which is close to the value of two predicted for the radiative lifetime ratio between Xs and charged excitons [34]. Now that the blinking "grey" ("on") level has been associated with the $X^+$s (Xs and $X^-$s), the appearance of the blinking "off" level could be caused by multiply-charged excitons or the exciton trapping process [25,26,28,35].

A possible origin of the fast lifetime $\tau_2$ from XXs can be safely ruled out based on the following discussions. First, the $X^+$ PL decay of the "grey" intensity level is dominated by Auger recombination with a fitted rate $K_{X+}$ of ~1/5.8 ns$^{-1}$ in Fig. 2(e). Meanwhile, the $X^-$ PL decay of the "on" intensity level is dominated by radiative recombination so that its Auger rate $K_{X-}$ should approach zero. Then the PL decay rate of a XX can be expressed as $K_{XX} = 2(K_{X+} + K_{X-}) \approx 2K_{X+}$ [36] to yield a maximum PL lifetime of ~2.9 ns that is much shorter than the $\tau_2$ value of ~33.4 ns. In fact, a XX Auger lifetime of ~90 ps was previously reported for ensemble CsPbI$_3$ NCs with similar sizes to the ones studied here [21]. Second, as shown in Fig. 2(g), the average area ratio between the central and the side $g^{(2)}(\tau)$ peaks was calculated to be ~0.08 (~0.04) for this single NC excited at $<N>$ = ~0.02 (~1.7), implying that the PL efficiency of XXs is at most ~4% relative to that of Xs [21,37,38]. This extremely-low PL efficiency of XXs contradicts the fact that the fast lifetime component contributes ~40% of the total photons in the "on" intensity level from the calculation of $(A_2\tau_2)/(A_1\tau_1 + A_2\tau_2)$[39]. It should be noted that, although we focused on a specific CsPbI$_3$ NC in Fig. 2, similar appearances of two charged-exciton species in the blinking "on" and "grey" intensity levels were observed in all of the ~20 single CsPbI$_3$ NCs excited in our experiment at high power (see Fig. S7 of the Supplemental Material [24] for another representative NC).

After room-temperature characterizations, we switched to the cryogenic temperature of ~4 K to measure the optical properties of single CsPbI$_3$ NCs excited first at a low excitation power corresponding to $<N>$ = ~0.1. As shown in Fig. 3(a), all of the single CsPbI$_3$ NCs studied in our experiment possessed an ultra-narrow PL linewidth of ~200 μeV that is limited only by our system resolution. Moreover, both the PL blinking and the spectral diffusion effects are completely suppressed in single CsPbI$_3$ NCs, as can be seen in Fig. 3(b) from the time-dependent PL spectral image of a representative NC. PL linewdiths as narrow as hundreds of μeV were occasionally observed in single CdSe NCs at the cryogenic temperature [17,19], however, their stability over a long measurement time like the one shown in Fig. 3(b) has never been reported in literature. In Fig. 3(c), we plot the PL decay curve measured for a representative CsPbI$_3$ NC, which can be fitted by a single-exponential function with a radiative lifetime of ~1.02 ns. Similar to the case previously reported for single perovskite CsPbBr$_3$ and CsPb(Cl/Br)$_3$ NCs [22,23], no long-lifetime component of dark-exciton emission commonly observed in single CdSe NCs [40] was resolved here from the PL dynamics of single CsPbI$_3$ NCs.

Next in our experiment, we excited the same single CsPbI$_3$ NCs at both $<N>$ = ~0.1 and ~0.8 to explore how their optical properties evolve with the increasing excitation power at 4 K. As representatively shown in Fig. 4(a) for a single CsPbI$_3$ NC, there existed mainly a single PL peak from Xs at $<N>$ = ~0.1 and an additional one emerged at a lower energy after $<N>$ was increased to ~0.8. Although each of the two peaks is still associated with a narrow PL linewidth at a given time, the spectral diffusion effect is slightly induced at $<N>$ = ~0.8 especially for the X peak, as can be seen in Fig. 4(b) from the time-dependent PL spectral image. This can be explained by the formation of charged excitons with fluctuating local fields to disturb the exciton energy by means of the Stark effect [16,17]. We attribute the red-shifted PL peak to be from X$^-$s, whose PL photons are mixed with those from Xs in the blinking "on" intensity levels of the room-temperature PL time trace measured at $<N>$ = ~1.7 in Fig. 2(c). Auger recombination should be still efficient for both XXs and X$^+$s at the cryogenic temperature since no other PL peaks were resolved from single CsPbI$_3$ NCs excited at $<N>$ = ~0.8.

In Fig. 4(c), we plot the PL intensity of the X peak as a function of the laser excitation power, where a linear increase towards saturation can be observed due to the band-filling effect. Meanwhile, the PL intensity of the X$^-$ peak demonstrates a sublinear (instead of quadratic) dependence on the laser excitation power to exclude again its possible origin from XXs [41]. As shown in Fig. 4(d), the PL decay curve measured at $<N>$ = ~0.8 for the X peak can be fitted with a single-exponential lifetime of ~1.08 ns, which is close to the one of ~1.04 ns measured at $<N>$ = ~0.1 (see Fig. S8 of the Supplemental Material [24]). This independence of the X radiative lifetime on the laser excitation power strongly confirms that the shortened PL lifetime of "on"-level photons observed at room temperature in Fig. 2(e) should be caused by X$^-$s created with the increase of $<N>$ from ~0.02 to ~1.7. The PL decay curve measured at $<N>$ = ~0.8 for the X$^-$ peak is also plotted in Fig. 4(d) and fitted with a single-exponential function to yield a PL lifetime of ~0.42 ns, from which a X/X$^-$ lifetime ratio of ~2.57 can be obtained. From statistical measurements on ~10 single CsPbI$_3$ NCs, an average X/X$^-$ lifetime ratio of ~2.38 was obtained that is comparable to the one of ~1.83 measured at room temperature in Fig. S6 of the Supplemental Material [24]. In Fig. 4(e), we plot a histogram of energy separations between the X and X$^-$ PL peaks measured over 15 single CsPbI$_3$ NCs and the average value of ~8.1 meV roughly reflects the binding energy between a single exciton and an extra electron with an attractive interaction.

To summarize, we have demonstrated slow Auger recombination of charged excitons and nonblinking PL behaviour in single CsPbI$_3$ NCs at both room and cryogenic temperatures. Moreover, a resolution-limited PL linewidth of ~200 μeV without the influence of spectral diffusion was measured for single $CsPbI_3$ NCs at the cryogenic temperature. A combination of the above optical properties has been actively pursued in previous optical studies of single metal-chalcogenide NCs without any success but is easily and repeatedly realized here in single $CsPbI_3$ NCs from many batches of samples synthesized by us with a facile colloidal approach. The above findings mark the emergence of a potent semiconductor nanostructure that will surely stimulate intensive research efforts in both fundamental studies and practical applications. We attribute the suppressed PL blinking and spectral diffusion effects in single $CsPbI_3$ NCs to the nonexistence of defect traps for band-edge charge carriers to form charged excitons. However, the sign assignments for the two types of charged excitons and the exact origins for their slightly- and strongly-reduced Auger recombinations are yet to be determined from future experimental and theoretical works. We are optimistic that, with high-resolution spectroscopic technique being further employed on a single $CsPbI_3$ NC, fine exciton structures could be resolved from the X PL peak. Its long-term stability without spectral diffusion has also provided a suitable platform to realize a variety of coherent optical measurements such as quantum interference and resonantly-excited fluorescence.

This work is supported by the National Basic Research Program of China (No. 2012CB921801), the National Natural Science Foundation of China (Nos. 11574147, 91321105, 11274161 and 11321063), Jiangsu Provincial Funds for Distinguished Young Scientists (No. BK20130012), and the PAPD of Jiangsu Higher Education Institutions. F.H and C.Y. contributed equally to this work.

*corresponding author.

wxiaoyong@nju.edu.cn, yuzhang@jlu.edu.cn, mxiao@uark.edu

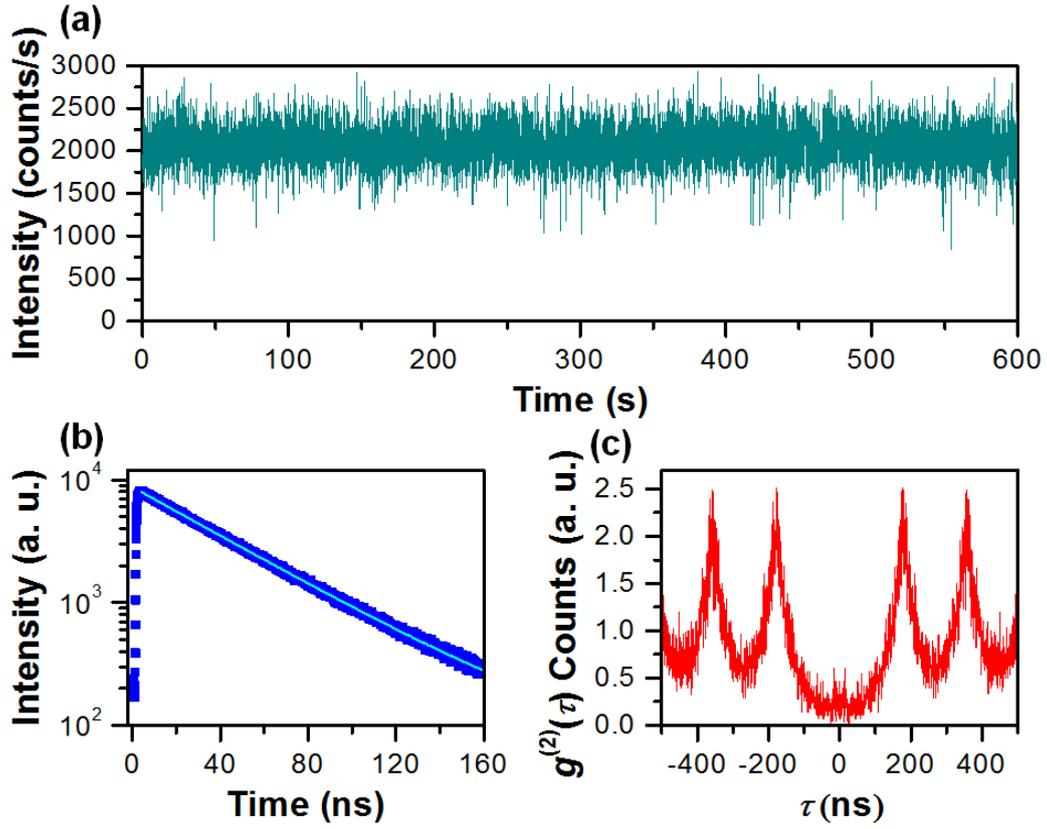

**FIG. 1. (a)** PL intensity time trace plotted with a binning time of 30 ms. **(b)** "On"-level PL decay curve fitted with a single-exponential lifetime of ~43.4 ns. **(c)** Second-order autocorrelation function $g^{(2)}(\tau)$ measurement. The above measurements were performed at room temperature for the same single $CsPbI_3$ NC excited at $<N> =$ ~0.03.

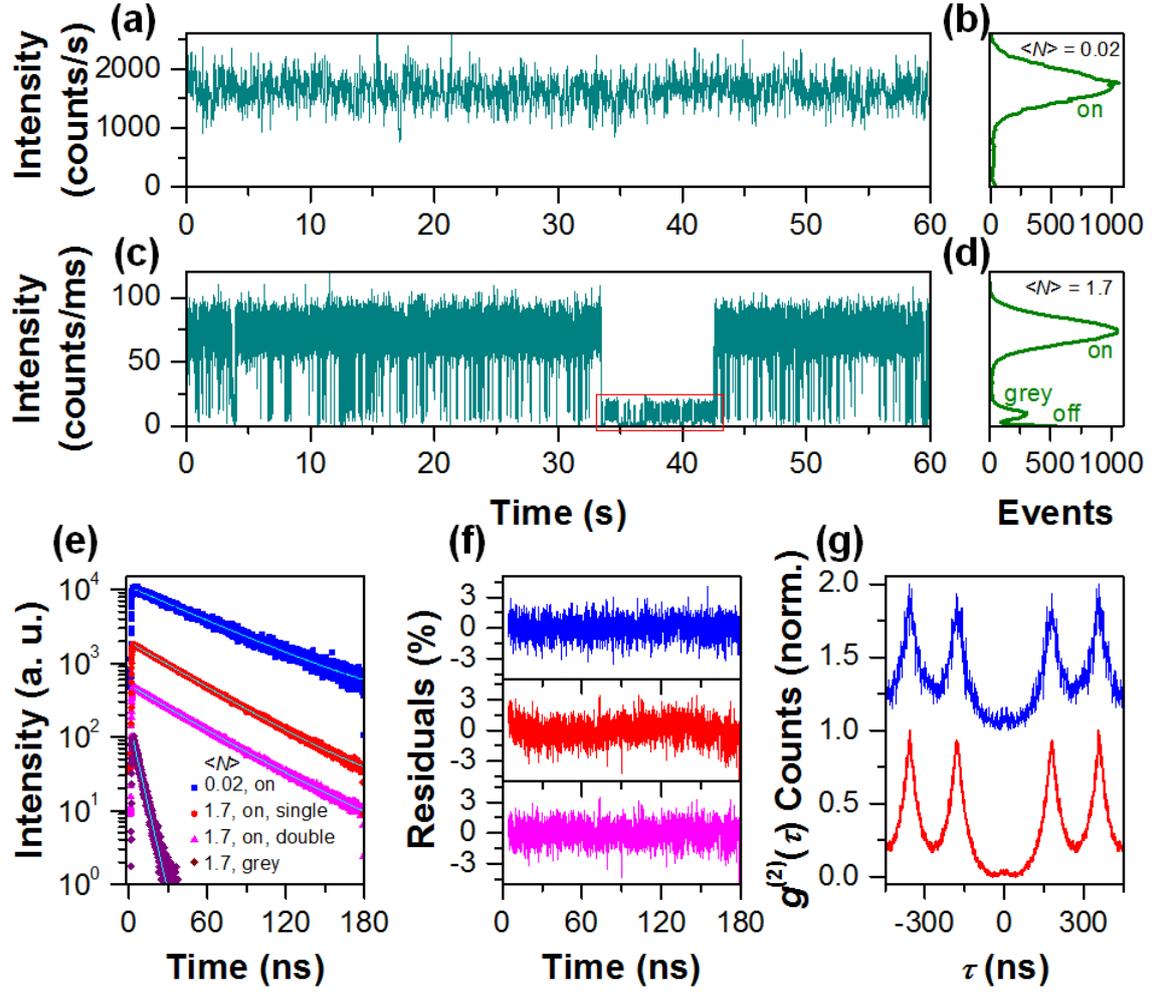

**FIG. 2.** (a) PL intensity time trace and (b) PL intensity distribution plotted with a binning time of 30 ms for a single CsPbI$_3$ NC excited at $<N>$ = ~0.02. (c) PL intensity time trace and (d) PL intensity distribution plotted with a binning time of 10 ms for the same single CsPbI$_3$ NC excited at $<N>$ = ~1.7. (e) "On"- or "grey"-level PL decay curve measured for the single CsPbI$_3$ NC excited at $<N>$ = ~0.02 or ~1.7. See descriptions in the figure and the text for measurement conditions and fitting details. (f) (from top to bottom) Fitting residuals for the first three PL decay curves plotted in (e). (g) Second-order autocorrelation function $g^{(2)}(\tau)$ measurements for the single CsPbI$_3$ NC excited at $<N>$ = ~0.02 (top) and ~1.7 (bottom), respectively. The above measurements were performed at room temperature.

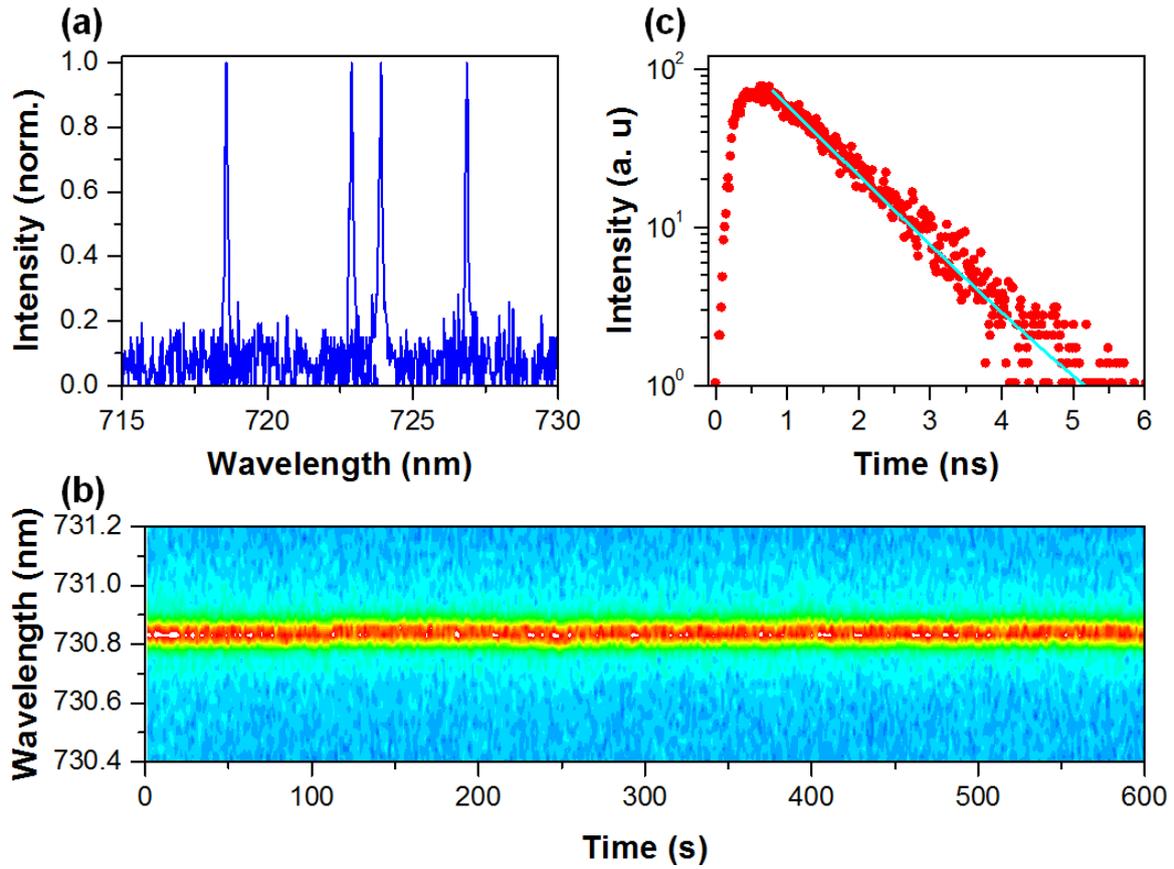

**FIG. 3. (a)** PL spectra of four single CsPbI$_3$ NCs with ultra-narrow linewidths of ~200 μeV. **(b)** Time-dependent PL spectral image of a single CsPbI$_3$ NC. **(c)** PL decay curve fitted with a single-exponential lifetime of ~1.02 ns for a single CsPbI$_3$ NC. The above measurements were performed at 4 K with the excitation of <*N*> = ~0.1.

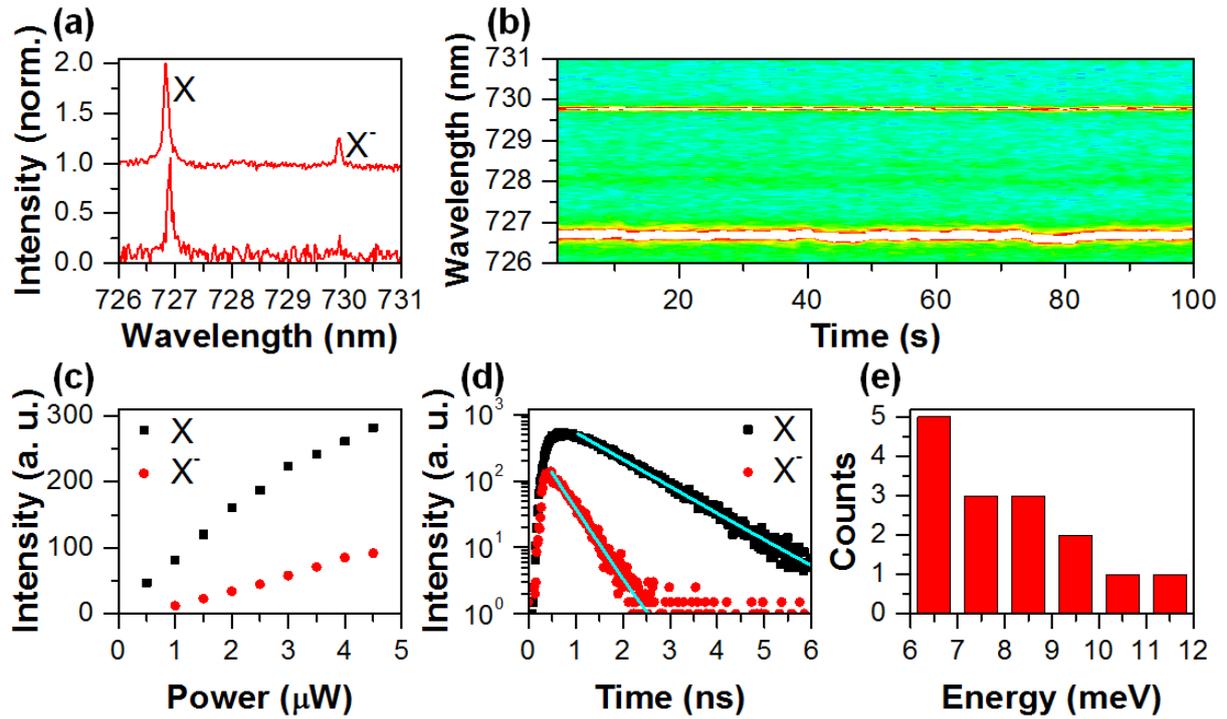

**FIG. 4.** (a) PL spectrum of a single $CsPbI_3$ NC excited at $<N>$ = ~0.1 with a dominant X peak (bottom) and PL spectrum of the same $CsPbI_3$ NC excited at $<N>$ = ~0.8 with the emergence of an additional $X^-$ peak (top). (b) Time-dependent PL spectral image of the single $CsPbI_3$ NC excited at $<N>$ = ~0.8 where the spectral diffusion effect is slightly manifested. The integration time for each PL data point is 1 s. (c) PL intensities of the X and $X^-$ peaks plotted as a function of the laser excitation power. (d) PL decay curves of the X and $X^-$ peaks each fitted with a single-exponential function. (e) Distribution of the energy separations between the X and $X^-$ PL peaks for ~15 single $CsPbI_3$ NCs. The above measurements were performed at 4 K.

# Supplemental Material

**Slow Auger Recombination of Charged Excitons in Nonblinking Perovskite Nanocrystals without Spectral Diffusion**

Fengrui Hu,[1,†] Chunyang Yin,[1,†] Huichao Zhang,[1,3] Chun Sun,[2] William W. Yu,[2] Chunfeng Zhang,[1] Xiaoyong Wang,[1,*] Yu Zhang,[2,*] and Min Xiao[1,4,*]

[1] *National Laboratory of Solid State Microstructures, School of Physics, and Collaborative Innovation Center of Advanced Microstructures, Nanjing University, Nanjing 210093, China*

[2] *State Key Laboratory on Integrated Optoelectronics and College of Electronic Science and Engineering, Jilin University, Changchun 130012, China*

[3] *College of Electronics and Information, Hangzhou Dianzi University, Xiasha Campus, Hangzhou 310018, China*

[4] *Department of Physics, University of Arkansas, Fayetteville, Arkansas 72701, USA*

[†]These authors contributed equally to this work.

[*]Correspondence to X. W. (wxiaoyong@nju.edu.cn), Y. Z. (yuzhang@jlu.edu.cn), and M. X. (mxiao@uark.edu).

**Experimental details**

**Synthesis of colloidal CsPbI$_3$ NCs.** For the preparation of Cs-oleate precursor, a mixture of Cs$_2$CO$_3$ (0.8 g), oleic acid (OA, 2.5 mL) and octadecene (ODE, 30 mL) was loaded into a 100 mL 3-neck flask and dried under vacuum for 1 h at 120 °C. The resulting Cs-oleate solution was first elevated to 150 °C until all the Cs$_2$CO$_3$ had been dissolved and then kept at 100 °C for the subsequent injection process. For the synthesis of CsPbI$_3$ NCs, a mixture of PbI$_2$ (0.174 g) and ODE (10 mL) was loaded into a 50 mL 3-neck flask and dried under vacuum for 1 h at 120 °C. After quick injection of OA (0.5 mL) and oleylamine (OLA, 0.5 mL) under N$_2$, the solution temperature was elevated to 160 °C for the reaction with the Cs-oleate precursor (0.8 mL). The reaction mixture was kept at this temperature for 5 s and then cooled down to room temperature by an ice-water bath. Finally, the CsPbI$_3$ NCs were purified by high-speed centrifugation (10000 rpm, 10 min) and then re-dispersed in toluene to form a stable solution. The solution absorption and emission spectra of ensemble CsPbI$_3$ NCs were measured with a V-550 UV-Visible spectrophotometer from Jasco, and the fluorescent quantum yield was estimated to be ~52%.

**Optical characterizations of single CsPbI$_3$ NCs.** One drop of the diluted solution of CsPbI$_3$ NCs with dissolved polymer of poly-DL-lactide was spin-coated onto a fused silica substrate to form a solid film for the single-NC optical characterizations. The 570 nm output of a 5.6 MHz, picosecond supercontinuum fiber laser (EXR-15, from NKT Photonics) was used as the excitation source. For the room-temperature measurements, the laser beam was focused onto the sample substrate by an immersion-oil objective (N.A. = 1.4). PL signal of a single NC was collected by the same objective and sent through a 0.5 m spectrometer to a charge-coupled-device camera for the PL spectral measurements. PL signal of a single NC could be alternatively sent through a non-polarizing 50/50 beam splitter to two avalanche photodiodes (APDs) in a time-correlated single-photon counting (TCSPC) system with a time resolution of ~250 ps. The TCSPC system was operated under the time-tagged, time-resolved mode so that the arrival times of each photon relative to the laboratory time and the laser pulse time could be both obtained, which allowed us to plot the PL intensity time trace and the PL decay curve, respectively. Moreover, the delay times between photons collected by one APD and those by the other could be summed up to yield the second-order autocorrelation $g^{(2)}(\tau)$ functions. For the optical characterizations of single $CsPbI_3$ NCs at cryogenic temperature, the sample substrate was contained in a helium-free cryostat operated at ~4 K. Very similar optical setups to those described above at room temperature were employed except that the immersion-oil objective was replaced by a dry objective (N.A. = 0.8).

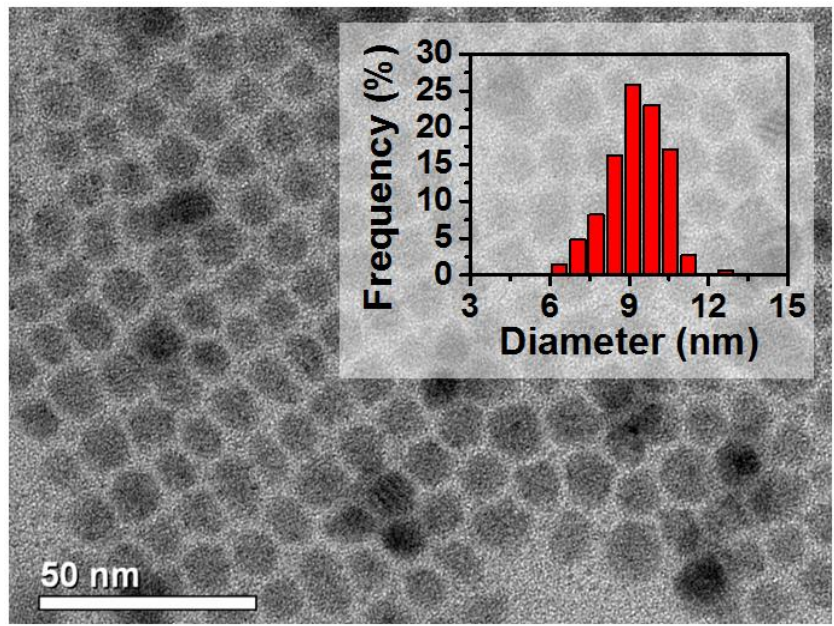

**Figure S1.** Transmission electron microscopy (TEM) image of $CsPbI_3$ NCs. Inset: Histogram for the size distribution of $CsPbI_3$ NCs.

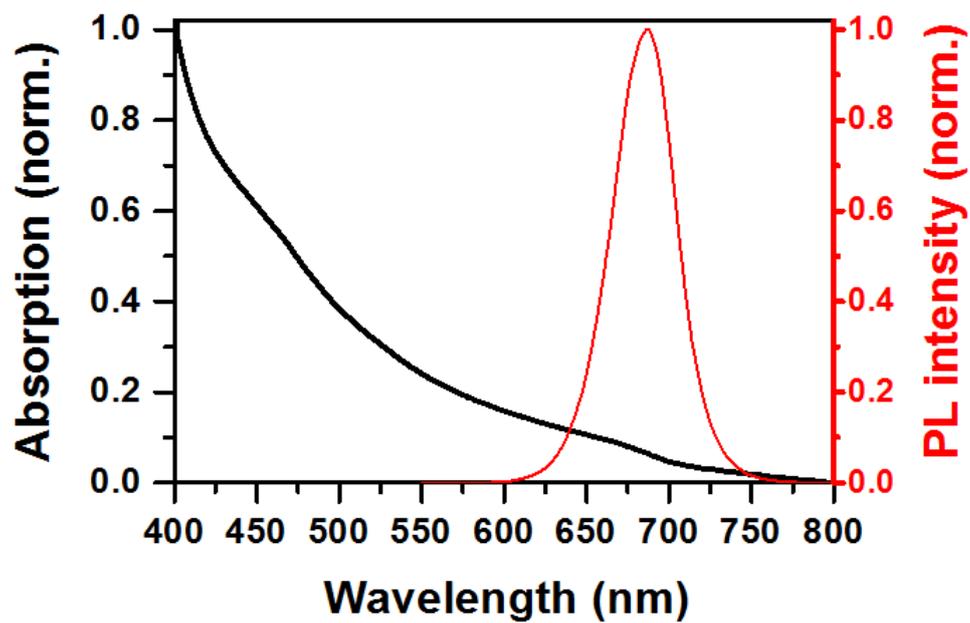

**Figure S2.** Solution absorption and emission spectra measured for ensemble CsPbI$_3$ NCs.

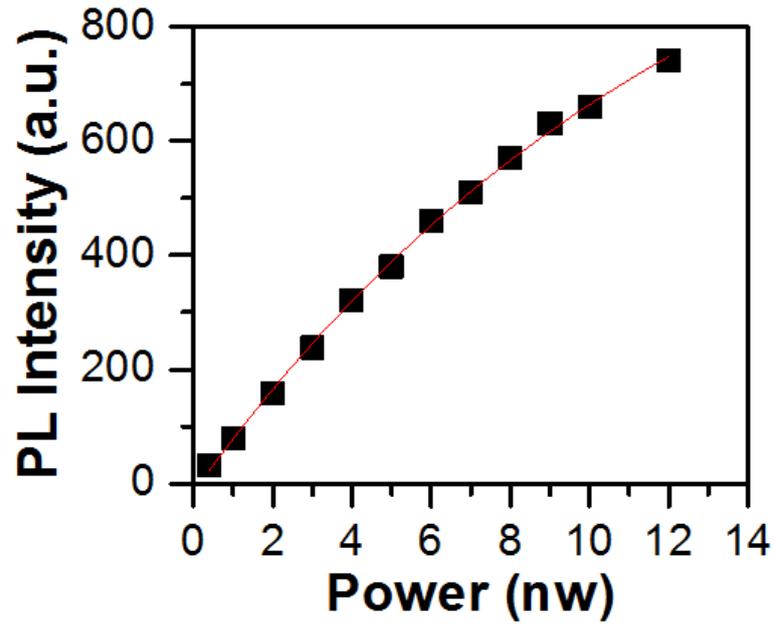

**Figure S3.** "On"-level PL intensity of a single CsPbI$_3$ NC measured as a function of the laser power $P$ and fitted with the function, $\propto 1 - e^{-\alpha P} = 1 - e^{-<N>}$, from which a fitting constant $\alpha$ can be obtained. The exciton number $<N>$ corresponding to a specific $P$ can thus be calculated from $<N> = \alpha P$.

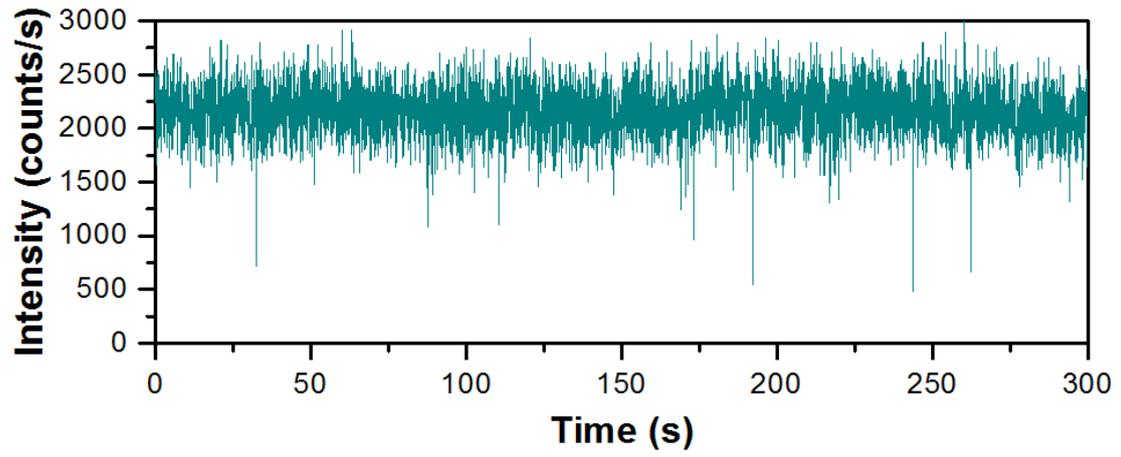

**Figure S4.** PL intensity time trace of a single CsPbI$_3$ NC excited at $\langle N \rangle$ = ~0.03, where occasional switchings of the PL intensity from the "on" to the lower intensity levels can be observed.

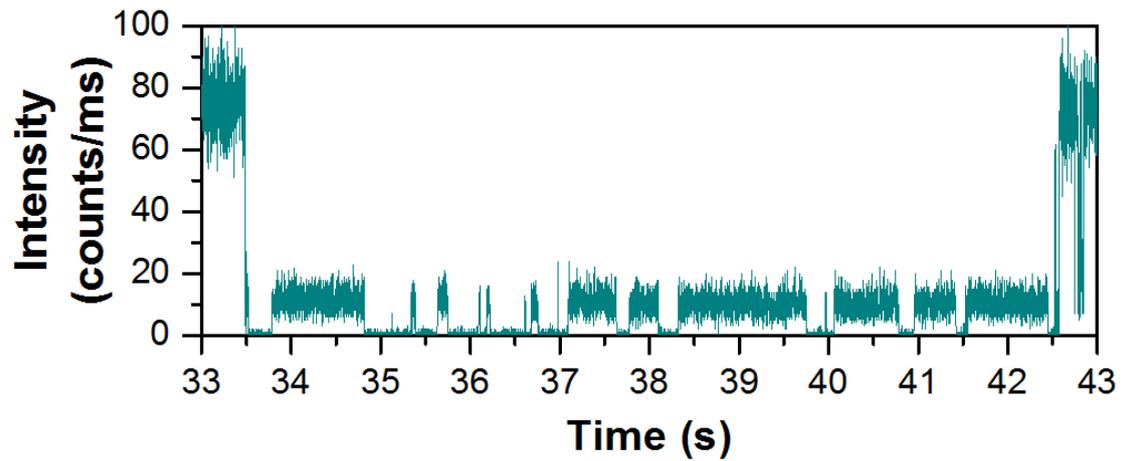

**Figure S5.** PL intensity time trace of a single CsPbI$_3$ NC excited at $<N>$ = ~1.7, where the blinking "on", "grey" and "off" levels can be clearly resolved. This is an enlarged plot for the part of the PL intensity time trace marked by a red box in Fig. 2(c).

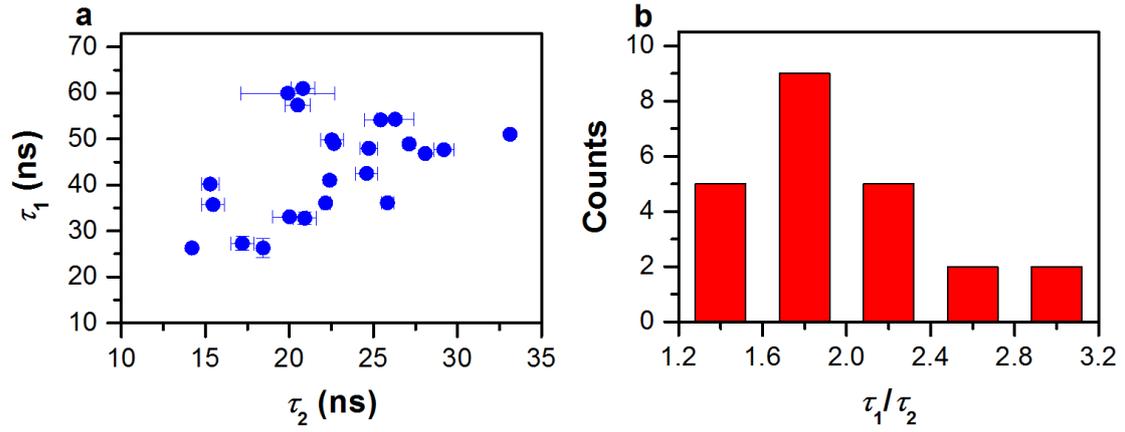

**Figure S6. (a)** The slow lifetime $\tau_1$ plotted as a function of the fast liftime $\tau_2$ for ~20 single CsPbI$_3$ NCs excited at $\langle N \rangle$ = ~1.7 at room temperature. **(b)** Distribution of the $\tau_1/\tau_2$ ratios for these ~20 single CsPbI$_3$ NCs.

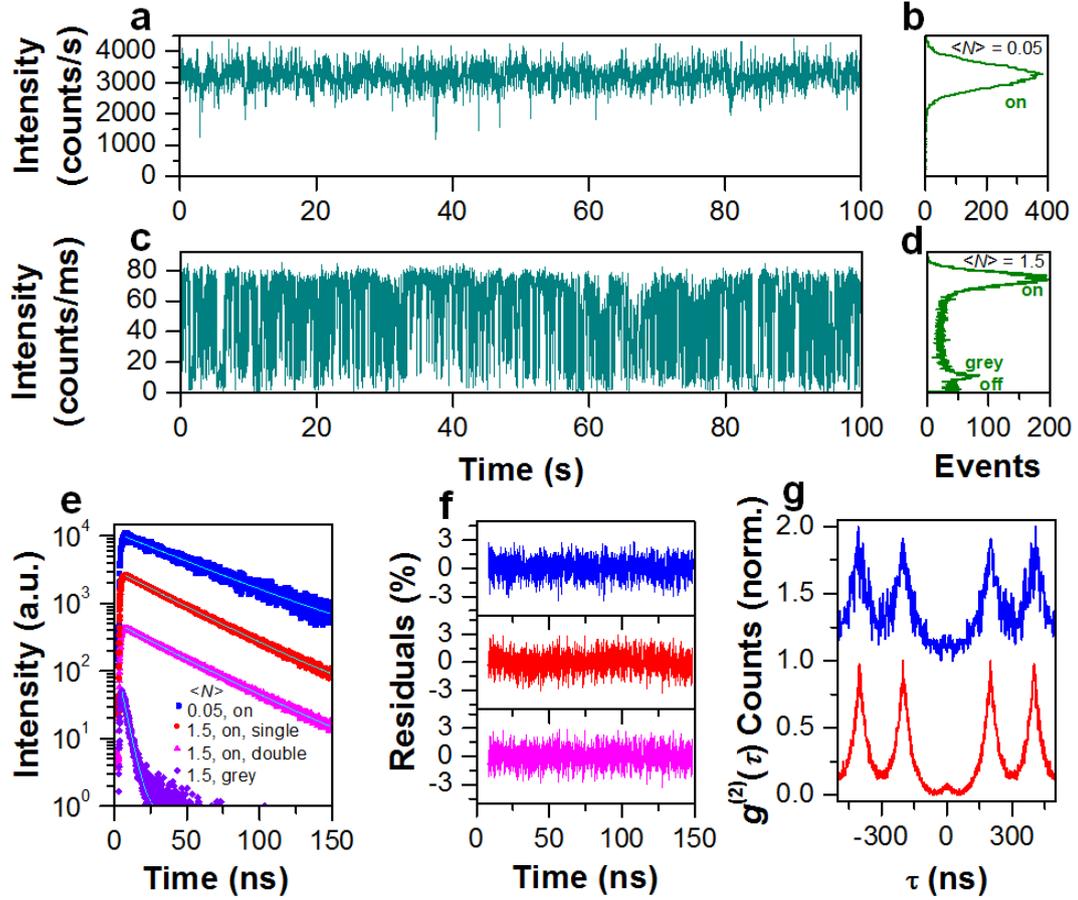

**Figure S7.** Room-temperature optical properties of a single CsPbI$_3$ NC excited at both low and high powers. (**a**) PL intensity time trace and (**b**) PL intensity distribution plotted with a binning time of 30 ms for a single CsPbI$_3$ NC excited at $<N>$ = ~0.05. (**c**) PL intensity time trace and (**d**) PL intensity distribution plotted with a binning time of 10 ms for the same single CsPbI$_3$ NC excited at $<N>$ = ~1.5. (**e**) (from top to bottom) "On"-level PL decay curve measured for the single CsPbI$_3$ NC excited at $<N>$ = ~0.05 and fitted with a single-exponential lifetime of ~49.7 ns. "On"-level PL decay curve measured for the single CsPbI$_3$ NC excited at $<N>$ = ~1.5 and fitted with a single-exponential lifetime of ~39.4 ns. "On"-level PL decay curve measured for the single CsPbI$_3$ NC excited at $<N>$ = ~1.5 and fitted by a double-exponential function with a slow and a fast lifetime of ~45.6 ns and ~30.2 ns, respectively. "Grey"-level PL decay curve measured for the single CsPbI$_3$ NC excited at $<N>$ = ~1.5 and fitted with a single-exponential lifetime of ~4.2 ns. (**f**) (from top to bottom) Fitting residuals for the first three PL decay curves plotted in (e). (**g**) Second-order autocorrelation function $g^{(2)}(\tau)$ measurements for the single CsPbI$_3$ NC excited at $<N>$ = ~0.05 (top) and ~1.5 (bottom), respectively.

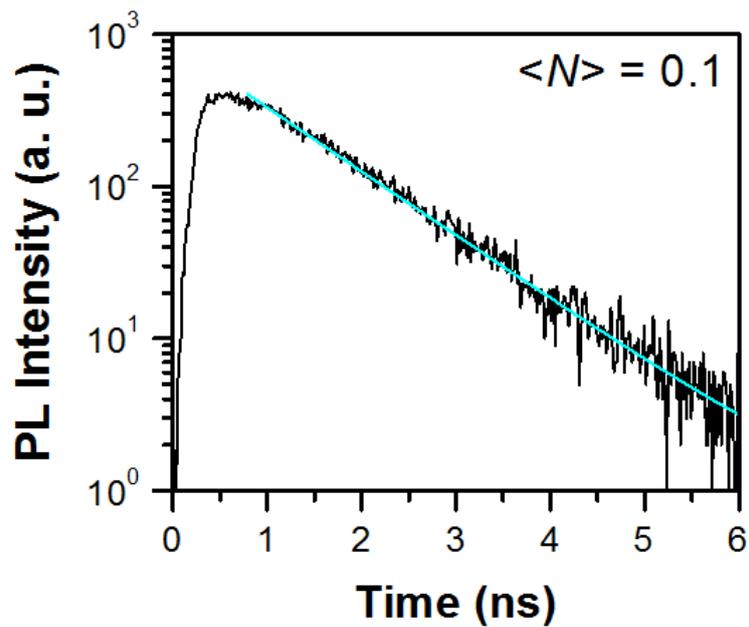

**Figure S8.** PL decay curve of neutral single excitons for a single CsPbI$_3$ NC excited at 4 K with <*N*> = ~0.1, which is fitted with a single-exponential lifetime of ~1.02 ns. PL decay curve of neutral single excitons for the same CsPbI$_3$ NC excited at 4 K with <*N*> = ~0.8 is plotted in Fig. 4(d) and fitted with a single-exponential lifetime of ~1.08 ns.